\documentstyle[12pt,graphicx]{article}
\begin{document}
\title{Comment to "Fine structure constant in the spacetime of a cosmic string}
\author{E. R. Bezerra de Mello \thanks{E-mail: emello@fisica.ufpb.br}\\
Departamento de F\'{\i}sica-CCEN\\
Universidade Federal da Para\'{\i}ba\\
58.059-970, J. Pessoa, PB\\
C. Postal 5.008\\
Brazil}
\maketitle       
\begin{abstract} 
In the paper {\it Phys. Lett. B {\bf 614} (2005) 140-142}, F. Nasseri shows that the values of the fine structure constant reduces due to the presence of a cosmic string. In this comment I want to point out that this conclusion is not completely correct in the sense that the result obtained is valid only in a very special situation.
\vspace{1pc}
\end{abstract}

\maketitle

In a recent paper F. Nasseri \cite{FN} analysed the Bohr's atom in the cosmic string spacetime. The author claims that the fine structure constant in this system reduces by a factor of order $8.736\times 10^{-17}$, consequently very small as the author says. In order to reach this conclusion the author considers separately two distinct forces acting on the electron: $i)$ the induced electrostatic self-force on the electron due to the presence of a cosmic string \cite{L,S}, and $ii)$ the electrostatic one between the electron and the proton. So the total force will be given by the sum of these two forces. Although being not explicitly said, the author assumes that the proton is placed on the cosmic string considered as an ideal one. The point that I want to call attention is to the fact that the induced force on the electron, Eq. $(11)$ of \cite{FN}, is perpendicular to the cosmic string assumed along the $z-$axis. So the correct expression to it is:
\begin{eqnarray}
{\vec{F}}=\frac{k\pi G\mu e^2}{4\rho^2}{\hat{\rho}} \ .
\end{eqnarray}
$\rho$ being the distance between the electron and the string. Also I have included the factor $k$ in the numerator of the right hand side of the above equation, missed in the Eq. $(11)$ of \cite{FN}\footnote{In Ref. \cite{FN} the author has  used the mks units where $k=1/(4\pi\epsilon_0)$.}. 

As to the electrostatic force, it is along the radial distance between the electron and the proton. Considering the origin of the referential system on the proton, this force  reads:
\begin{eqnarray}
{\vec{F}}=-\frac{ke^2}{r^2}{\hat{r}} \ .	
\end{eqnarray}
So in principle both forces are not always at the same direction. 

The Eq. $(15)$ of \cite{FN} is correct only when electron and proton are in the plane orthogonal to the string. Consequently the correction on the fine structure constant presented in Eq. $(17)$ is not general.\\
\noindent
{\bf Acknowledgment}\\
\noindent
I want to thanks the referee for pertinent comments on this Comment. This work was partially supported by CNPq.

\end{document}